\documentclass[fleqn]{llncs}
\newcommand{\eclipse}{ECL$^i$PS$^e$}
\newcommand{\oldbfit}[1]{\begin{bfseries}\textit{#1}\end{bfseries}}

\usepackage{amsfonts, amssymb, latexsym, graphicx}

\usepackage{pstricks}
\usepackage{fancybox}
\usepackage{times}
\usepackage{latexsym}

\newcommand{\oldbfe}[1]{\begin{bfseries}\emph{#1}\end{bfseries}}

\newcommand{\ES}{\mbox{$\emptyset$}}

\newcommand{\myra}{\mbox{$\:\rightarrow\:$}}

\newcommand{\A}{\mbox{$\ \wedge\ $}}

\newcommand{\Orr}{\mbox{$\ \vee\ $}}

\newcommand{\sse}{\mbox{$\:\subseteq\:$}}

\newcommand{\po}{\mbox{$\ \sqsubseteq\ $}}

\newcommand{\fa}{\mbox{$\forall$}}
\newcommand{\te}{\mbox{$\exists$}}

\newcommand{\LL}{\mbox{$\ldots$}}

\newcommand{\C}[1]{\mbox{$\{{#1}\}$}}           

\newcommand{\NI}{\noindent}
\newcommand{\HB}{\hfill{$\Box$}}

\newcommand{\III}{\vspace{3 mm}}
\newcommand{\II}{\vspace{2 mm}}

\newcommand{\szkew}[1]{\relax \setbox0=\hbox{\kern -24pt $\displaystyle#1$\kern 0pt }%
\box0}
{\catcode`\@=11 \global\let\ifjusthvtest@=\iffalse}

\newcounter{oldmycaption}

\newcommand{\mycite}[1]{\cite{#1}\glossary{#1}}

\newcommand{\p}[2]{\langle #1 \ ; \ #2 \rangle}

\usepackage{amsmath}
\usepackage{colordvi}
\usepackage{latexsym}

\title{Explaining Constraint Programming}

\author{Krzysztof R. Apt\inst{1,2,3}}
\institute{School of Computing, National University of Singapore
\and
CWI, Amsterdam
\and
University of Amsterdam, the Netherlands
}
\date{}

\begin{document}

\maketitle

\begin{abstract}

  We discuss here constraint programming (CP) by using a
  proof-theore\-tic perspective.  To this end we identify
  three levels of abstraction. Each level sheds light on the essence
  of CP.

In particular, the highest level allows us to bring CP closer to
the computation as deduction paradigm.  At the middle level we can
explain various constraint propagation algorithms. Finally, at the
lowest level we can address the issue of automatic generation and
optimization of the constraint propagation algorithms.
\end{abstract}
\section{Introduction}

Constraint programming is an alternative approach to programming
which consists of modelling the problem as a set of requirements
(constraints) that are subsequently solved by means of general and domain specific
methods.

Historically, constraint programming is an outcome of a long process
that has started in the seventies, when the seminal works of Waltz and
others on computer vision (see, e.g.,\cite{Waltz75})
led to identification of constraint
satisfaction problems as an area of Artificial Intelligence.  In this
area several fundamental techniques, including constraint propagation
and enhanced forms of search have been developed.

In the eighties, starting with the seminal works of Colmerauer (see,
e.g., \cite{COLMERAUER87}) and Jaffar and Lassez (see
\cite{jaffar-constraint-87}) the area constraint logic programming was
founded.  In the nineties a number of alternative approaches to
constraint programming were realized, in particular in ILOG solver,
see e.g., \cite{ILOG03}, that is based on modeling the constraint
satisfaction problems in C++ using classes.  Another, recent, example
is the Koalog Constraint Solver, see \cite{Koa05}, realized as a Java
library.

This way constraint programming eventually emerged as a distinctive
approach to programming.  In this paper we try to clarify this
programming style and to assess it using a proof-theoretic perspective
considered at various levels of abstraction.  We believe that this
presentation of constraint programming allows us to more easily compare it 
with other programming styles and to isolate its salient features.

\section{Preliminaries}

Let us start by introducing 
the already mentioned concept of a constraint satisfaction problem.
Consider a sequence $X = x_1, \LL, x_m$ of variables with respective
domains $D_1, \LL, D_n$.  By a \oldbfe{constraint} on $X$ we mean
a subset of $D_1 \times \LL \times D_m$.  A \oldbfe{constraint
  satisfaction problem (CSP)} consists of a finite sequence of
variables $x_1, \LL, x_n$ with respective domains $D_1, \LL, D_n$
and a finite set ${\cal C}$ of
constraints, each on a subsequence of $X$.  
We write such a CSP as
\[
\p{{\cal C}}{x_1 \in D_1, \LL, x_n \in D_n}.
\]

A \oldbfe{solution} to a CSP is an assignment of values to its
variables from their domains that satisfies all constraints.  We say
that a CSP is \oldbfe{consistent} if it has a solution,
\oldbfe{solved} if each assignment is a solution, and
\oldbfe{failed} if either a variable domain is empty or a constraint
is empty.  Intuitively, a failed CSP is one that obviously does not
have any solution.  In contrast, it is not obvious at all to verify
whether a CSP is solved. So we introduce an imprecise concept of a
`\oldbfe{manifestly solved}' CSP which means that it is computationally
straightforward to verify that the CSP is solved. So this notion depends on
what we assume as `computationally straightforward'.

In practice the constraints are
written in a first-order language.  They are then atomic formulas or
simple combinations of atomic formulas. One identifies then a
constraint with its syntactic description.
In what follows we study CSPs with finite domains.  

\section{High Level}
\label{sec:high}

At the highest level of abstraction constraint programming can be seen
as a task of formulating specifications as a CSP and of solving it.
The most common approach to solving a CSP is based on a
\oldbfe{top-down search} combined with \oldbfe{constraint
  propagation}.  

The top-down search is determined by a
\emph{splitting strategy} that controls the splitting of a given CSP
into two or more CSPs, the `union' of which (defined in the natural
sense) is \oldbfe{equivalent} to (i.e, has the same solutions as) the
initial CSP.  In the most common form of splitting a variable is
selected and its domain is partitioned into two or more parts.  The
splitting strategy then determines which variable is to be selected
and how its domain is to be split.  

\begin{figure}[htbp]
  \centering
\input{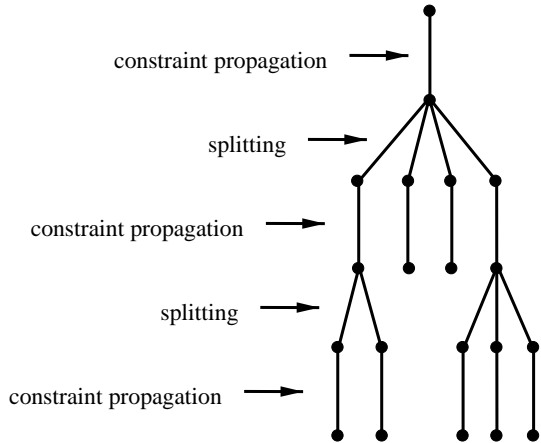}  
\begin{center}
\caption{\label{fig:search}A search tree for a CSP}  
\end{center}
\end{figure}

In turn, constraint propagation
transforms a given CSP into one that is equivalent but \emph{simpler},
i.e, easier to solve. Each form of constraint propagation determines
a notion of \oldbfe{local consistency} that in a loose sense approximates
the notion of consistency and is computationally efficient to achieve.
This process leads to a search tree in which constraint propagation is
alternated with splitting, see Figure \ref{fig:search}.

So the nodes in the tree are CSPs with the root (level 0) being the
original CSP.  At the even levels the constraint propagation is
applied to the current CSP.  This yields exactly one direct
descendant.  At the odd levels splitting is applied to the current
CSP.  This yields more than one descendant.  The leaves of the tree
are CSPs that are either failed or manifestly solved.  So from the
leaves of the trees it is straightforward to collect all the solutions
to the original CSP.

The process of tree generation can be expressed by means of proof rules
that are used to express transformations of CSPs.  In general we have two types
of rules. The \oldbfe{deterministic} rules transform a given CSP into
another one. We write such a rule as:
\[
\frac{\phi}
     {\psi}
\]
where $\phi$ and $\psi$ are CSPs. 

In turn, the \oldbfe{splitting} rules transform a given CSP into a sequence of CSPs.
We write such a rule as:
\[
\frac{\phi}
     {\psi_1 \mid \LL \mid \psi_n}
\]
where $\phi$ and $\psi_1, \LL, \psi_n$ are CSPs. 

It is now easy to define the notion of an \oldbfe{application of a proof rule} to
a CSP. In the case of a deterministic rule we just replace (after an appropriate renaming)
the part that matches the premise of the rule by the conclusion.
In the case of a splitting rule we replace (again after an appropriate renaming)
the part that matches the premise of the rule by \emph{one} of the CSPs $\psi_i$ from the 
rule conclusion.

We now say that 
a deterministic rule
\[
\frac{\phi}{\psi}
\]
is \oldbfe{equivalence preserving} if $\phi$ and $\psi$ are equivalent
and that a splitting rule
\[
\frac{\phi}{\psi_1 \mid \LL \mid \psi_n}
\]
is \oldbfe{equivalence preserving} if the union
of $\psi_i$'s is equivalent to $\phi$.

In what follows all considered rules will be equivalence preserving.
In general, the deterministic rules are more `fine grained' than the constraint propagation
step that is modeled as a single `step' in the search tree.
In fact, our intention is to model constraint propagation as a
\emph{repeated} application of deterministic rules. In the next section we shall
discuss how to schedule these rule applications efficiently.

The search for solutions can now be described by means of
\oldbfe{derivations}, just like in logic programming.
In logic programming we have in general two types of finite derivations: successful and failed.
In the case of proof rules as defined above a new type of derivations naturally arises.

\begin{definition} \label{def:derivations}
Assume a finite set of proof rules.  

  \begin{itemize}
  \item By a \oldbfit{derivation} we mean a sequence of CSPs such that
    each of them is obtained from the previous one by an
    application of a proof rule.

  \item A finite derivation is called

    \begin{itemize}
    \item \oldbfit{successful}
if its last element is a first manifestly solved CSP in this derivation,
        
      \item \oldbfit{failed} 
if its last element is a first failed CSP in
        this derivation,
      
\item \oldbfit{stabilizing} 
if its last element is a first CSP in this derivation that is 
closed under the applications of the considered proof rules. 
\HB
    \end{itemize}

   \end{itemize}
\end{definition}

The search for a solution to a CSP can now be described as a search
for a successful derivation, much like in the case of logic
programming.  A new element is the presence of stabilizing
derivations.

One of the main problems constraint programming needs to deal with is
how to limit the size of a search tree.  At the high level of
abstraction this matter can be addressed by focusing on the
derivations in which the applications of splitting rules are postponed
as long as possible.  This bring us to a consideration of stabilizing
derivations that involve only deterministic rules.  In practice such
derivations are used to model the process of constraint propagation.
They do not lead to a manifestly solved CSP but only to a CSP that is
closed under the considered deterministic rules. So solving the
resulting CSP requires first an application of a splitting rule.  (The
resulting CSP can be solved but to determine it may be computationally
expensive.)

This discussion shows that at a high level of abstraction constraint
programming can be viewed as a realization of the \oldbfe{computation
  as deduction} paradigm according to which the computation process is
identified with a constructive proof of a formula from a set of
axioms.  In the case of constraint programming such a constructive
proof is a successful derivation. Each such derivation yields at least
one solution to the initial CSP.

Because so far no specific rules are considered not much more can be
said at this level.  However, this high level of abstraction allows us
to set the stage for more specific considerations that belong to the
middle level.

\section{Middle Level}
\label{sec:middle}

The middle level is concerned with the form of derivations
that involve only deterministic rules. It
allows us to explain the \oldbfe{constraint propagation
  algorithms} which are used to enforce constraint propagation.  In
our framework these algorithms are simply efficient schedulers of
appropriate deterministic rules.  To clarify this point we now
introduce examples of specific classes of deterministic rules.
In each case we discuss a scheduler that can be used to schedule the
considered rules.

\subsection*{Example 1: Domain Reduction Rules}

These are rules of the following form:

\[
\frac{\p{{\cal C}}{x_1 \in D_1, \LL, x_n \in D_n}}
     {\p{{\cal C'}}{x_1 \in D'_1, \LL, x_n \in D'_n}}
\]
where $D'_i \sse D_i$ for all $i \in [1..n]$ and ${\cal C'}$ is the result of
restricting each constraint in ${\cal C}$ to $D'_1, \LL,  D'_n$.
\II

We say that such a rule is \oldbfe{monotonic} if, when viewed as a function
$f$ from the original domains $D_1, \LL, D_n$ to the reduced
domains $D'_1, \LL, D'_n$, i.e., 
\[
f(D_1, \LL, D_n)$  $:= (D'_1, \LL, D'_n),
\]
it is monotonic:
\[
\mbox{$D_i \sse E_i$ for all $i \in [1..n]$ implies $f(D_1, \LL, D_n) \sse f(E_1, \LL, E_n)$.}
\]
That is, smaller variable domains yield smaller reduced domains.

Now, the following useful result shows that a large number of
domain reduction rules are monotonic.

\begin{theorem} (\cite{Apt03})
Suppose each $D'_i$
is obtained from $D_i$
using a combination of 

\begin{itemize}
\item union and intersection operations,
\item transposition and composition operations applied to binary relations,
\item join operation $\Join$,
\item projection functions, and
\item removal of an element.
\end{itemize}

Then the domain reduction rule is monotonic.
\end{theorem}

This repertoire of operations is sufficient to describe typical domain
reduction rules considered in various constraint solvers used in
constraint programming systems, including solvers for Boolean
constraints, linear constraints over integers, and arithmetic
constraints over reals, see, e.g., \cite{Apt03}.

Monotonic domain reduction rules are useful for two reasons. First, we have the following observation.

\begin{note} \label{note:monotonic}
  Assume a finite set of monotonic domain reduction rules and an
  initial CSP ${\cal P}$.  Every stabilizing derivation starting in
  ${\cal P}$ yields the same outcome.
\end{note}

Second, monotonic domain reduction rules can be scheduled more
efficiently than by means of a naive round-robin strategy. This is
achieved by using a \oldbfe{generic iteration algorithm} which in its
most general form computes the least common fixpoint of a set of
functions $F$ in an appropriate partial ordering.  This has been
observed in varying forms of generality in the works of \cite{Ben96},
\cite{TU96}, \cite{FFS98} and \cite{Apt99b}.  This algorithm has the
following form.  We assume here a finite set of functions $F$, each
operating on a given partial ordering with the least element $\bot$.
\newpage

\II

\NI 
{\sc Generic Iteration} algorithm
\begin{tabbing}
\= $d := \bot$; \\
\> $G := F$; \\ 
{\tt WHILE} $G \neq \ES$ {\tt DO} \\
\> \quad choose $g \in G$; \\
\> \quad {\tt IF} $d \neq g(d)$ {\tt THEN} \\
\> \quad \quad $G := G \cup update(G,g,d)$; \\
\> \quad \quad $d := g(d)$ \\
\> \quad {\tt ELSE} \\
\> \quad \quad $G := G - \C{g}$ \\
\> \quad {\tt END} \\
\> {\tt END} 
\end{tabbing} 
where for all $G,g,d$

\begin{description}
\item[A] $\C{f \in F - G  \mid f(d) = d \A f(g(d)) \neq g(d)} \sse update(G,g,d)$.
\end{description}
\III

The intuition behind the assumption {\bf A} is that $update(G,g,d)$
contains at least all the functions from $F - G$ for which $d$ is a
fixpoint but $g(d)$ is not.  So at each loop iteration if $d \neq
g(d)$, such functions are added to the set $G$.  Otherwise the
function $g$ is removed from $G$.

An obvious way to satisfy assumption {\bf A} is by using the following
\emph{update} function:
\[
\emph{update}(G,g,d) := \C{f \in F - G  \mid f(d) = d \A f(g(d)) \neq g(d)}.
\]
The problem with this 
choice of \emph{update} is that it is
expensive to compute because for each function $f$ in $F - G$ we would have to
compute the values $f(g(d))$ and $f(d)$. So in practice, we are
interested in some approximations from above of this \emph{update} function
that are easy to compute. 
We shall return to this matter in a moment.

First let us clarify the status of the above algorithm.  Recall that a
function $f$ on a partial ordering $(D, \po )$ is called
\oldbfe{monotonic} if $x \po y$ implies $f(x) \po f(y)$ for all $x, y$
and \oldbfe{inflationary} if $x \po f(x)$ for all $x$.

\begin{theorem}\label{thm:GI}(\cite{Apt99b})
  Suppose that $(D, \po )$ is a finite partial ordering with the least
  element $\bot$. Let $F$ be a finite set of monotonic and
  inflationary functions on $D$. Then every execution of the {\sc Generic Iteration}
  algorithm terminates and computes in $d$ the least common fixpoint
  of the functions from $F$.  
\end{theorem}

In the applications we study the iterations carried out on a partial
ordering that is a Cartesian product of the component partial orderings.
More precisely, given $n$ partial orderings $(D_i, \po_i)$, each with
the least element $\bot_{i}$, we assume that
each considered function $g$ is defined on a `partial' Cartesian
product $D_{i_1} \times \LL \times D_{i_l}$. Here $i_1, \LL, i_l$ is a
subsequence of $1, \LL, n$ that we call the \emph{scheme} of $g$. 
Given $d \in D_1 \times \cdots \times D_n$, where $d := d_1, \LL, d_n$, 
and a scheme $s := i_1, \LL, i_l$ we denote by $d[s]$ the sequence
$d_{i_1}, \LL, d_{i_l}$.

The corresponding instance of the above {\sc Generic Iteration}
algorithm then takes the following form.  
\newpage

\II

\NI
{\sc Generic Iteration for Compound Domains} algorithm
\begin{tabbing}
\= $d := (\bot_1, \LL, \bot_n)$; \\
\> $d' := d$; \\ 
\> $G := F$; \\ 
\> {\tt WHILE} $G \neq \ES$ {\tt DO} \\
\> \quad choose $g \in G$; \\
\> \quad $d'[s] := g(d[s])$, where $s$ is the scheme of $g$; \\
\> \quad {\tt IF} $d'[s] \neq d[s]$ {\tt THEN}\\
\> \quad \quad $G := G \cup \C{f \in  F  \mid \mbox{scheme of $f$ includes $i$ such that $d[i] \neq d'[i]$}}$; \\
\> \quad \quad $d[s] := d'[s]$ \\
\> \quad {\tt ELSE} \\
\> \quad \quad $G := G - \C{g}$ \\
\> \quad {\tt END} \\
\> {\tt END} 
\end{tabbing}
\III

So this algorithm uses an \emph{update} function that is
straightforward to compute. It simply checks which components of $d$
are modified and selects the functions that depend on these
components.  It is a standard scheduling algorithm used in most
constraint programming systems.

\subsection*{Example 2: Arc Consistency}

\oldbfe{Arc consistency}, introduced in \mycite{mackworth-consistency},
 is the most popular notion of local consistency considered in constraint programming. 
Let us recall the definition.

\begin{definition} \mbox{} \\[-6mm]
  \begin{itemize}
  \item 
Consider a binary constraint $C$ on the variables $x, y$ with the domains 
$D_x$ and $D_y$, that is $C \sse D_x \times D_y$. We call $C$
\oldbfit{arc consistent} if
\begin{itemize}

\item
$\fa a \in D_x \te b \in D_y \: (a,b) \in C$,

\item
$\fa b \in D_y \te a \in D_x \: (a,b) \in C$.

\end{itemize}
\item We call a CSP \oldbfit{arc consistent} \index{arc consistency}
\index{local consistency notions!arc consistency} 
\index{CSP!arc consistent|see{local consistency notions}} 
if all its binary constraints are arc consistent.
  \end{itemize}
\end{definition}

So a binary constraint is arc consistent if every value in each domain
has a \oldbfe{support} in the other domain, where we call $b$ a support
for $a$ if the pair $(a,b)$ (or, depending on the ordering of the
variables, $(b,a)$) belongs to the constraint.

In the literature several arc consistency algorithms have been
proposed.  Their purpose is to transform a given CSP into one that is
arc consistent without losing any solution.  We shall now illustrate
how the most popular arc consistency algorithm, \texttt{AC-3}, due to
\mycite{mackworth-consistency}, can be explained as a specific scheduling
of the appropriate domain reduction rules. First, let us define the notion
of arc consistency in terms of such rules.

Assume a binary constraint $C$ on the variables $x, y$. We introduce the following two rules.
\newpage

\II

\begin{center}
{\em ARC CONSISTENCY 1}    
\end{center}

\[
\frac
{\p{C}{x \in D_x, y \in D_y}}
{\p{C}{x \in D'_x, y \in D_y}}
\]
where
$D'_x := \C{a \in D_x \mid \te \: b \in D_y \: (a,b) \in C}$.

\II

\begin{center}
{\em ARC CONSISTENCY 2}    
\end{center}

\[
\frac
{\p{C}{x \in D_x, y \in D_y}}
{\p{C}{x \in D_x, y \in D'_y}}
\]
where
$D'_y := \C{b \in D_y \mid \te \: a \in D_x \: (a,b) \in C}$.
\II

So in each rule a selected variable domain is reduced by retaining only 
the supported values.
The following observation characterizes the notion of arc consistency
in terms of the above two rules.

\begin{note}[Arc Consistency] \label{not:arc}
A CSP is arc consistent iff
it is closed under the applications of the
{\it ARC CONSISTENCY} rules 1 and 2.
\end{note}

So to transform a given CSP into an equivalent one that is arc
consistent it suffices to repeatedly apply the above two rules for all
present binary constraints. Since these rules are monotonic, we can
schedule them using the {\sc Generic Iteration for Compound Domains}
algorithm. However, in the case of the above rules an \emph{improved}
generic iteration algorithm can be employed that takes into account
commutativity and idempotence of the considered functions, see
\cite{Apt00a}.

Recall that given two functions $f$ and $g$ on a partial ordering we say that $f$
is \oldbfe{idempotent} if $f (f(x)) = f(x)$ for all $x$ and say that $f$
and $g$ \oldbfe{commute} if $f (g (x)) = g (f (x))$ for all $x$.
The relevant observation concerning these two properties is the following.

\begin{note}
  Suppose that all functions in $F$ are idempotent and that for
  each function $g$ we have a set of functions $Comm(g)$ from $F$ such
  that each element of $Comm(g)$ commutes with $g$.  If
  $update(G,g,d)$ satisfies the assumption {\bf A}, then so does the
  function $update(G,g,d) - Comm(g)$.
\end{note}

In practice it means that in each iteration of the generic iteration
algorithm less functions need to be added to the set $G$.  This
yields a more efficient algorithm.

In the case of arc consistency for each binary constraint $C$ the
functions corresponding to the {\it ARC CONSISTENCY} rules 1 and 2
referring to $C$ commute. Also, given two binary constraints that
share the first (resp. second) variable, the corresponding {\it ARC
  CONSISTENCY} rules 1 (resp. 2) for these two constraints commute, as
well.  Further, all such functions are idempotent.  So, thanks to the
above Note, we can use an appropriately `tighter' \emph{update} function.  The
resulting algorithm is equivalent to the \texttt{AC-3} algorithm.

\subsection*{Example 3: Constructive Disjunction}

One of the main reasons for combinatorial explosion in search for
solutions to a CSP are \oldbfe{disjunctive constraints}.  A typical
example is the following constraint used in scheduling problems:
\[
\begin{array}{l}
\texttt{Start[task}_1] + \texttt{Duration[task}_1] \leq \texttt{Start[task}_2] \Orr \\
\texttt{Start[task}_2] + \texttt{Duration[task}_2] \leq \texttt{Start[task}_1]
\end{array}
\]
stating
that either $\texttt{task}_1$ is scheduled before $\texttt{task}_2$
or vice versa. To deal with a disjunctive constraint we can apply the following splitting rule
(we omit here the information about the variable domains):

\[
\frac{C_1 \Orr C_2}{C_1 \mid C_2}
\]
which amounts to a case analysis.

However, as already explained in Section \ref{sec:high}
it is in general preferable to postpone an application of a
splitting rule and try to reduce the domains first.
\oldbfe{Constructive disjunction}, see \cite{vHSD98}, is a technique
that occasionally allows us to do this.  It can be expressed in our
rule-based framework as a domain reduction rule that uses some
auxiliary derivations as side conditions:
\III

\NI
\begin{center}
{\em CONSTRUCTIVE DISJUNCTION}  
\end{center}

\[
{\begin{array}{l}
\p{C_1 \Orr C_2}{x_1 \in D_1 , \LL, x_n \in D_n} \\
[-\medskipamount]
\hrulefill                                                      \\
\p{C'_1 \Orr C'_2}{x_1 \in D'_1 \cup D''_1, \LL, x_n \in D'_n \cup D''_n}
\end{array}
}
\mbox{\hspace{5mm}  where \hspace{5mm} $der_1, der_2$}
\]

\NI
with
\II

\NI
$der_1 := \p{C_1}{x_1 \in D_1, \LL, x_n \in D_n} \vdash \p{C'_1}{x_1 \in D'_1, \LL, x_n \in D'_n}$,

\NI
$der_2 := \p{C_2}{x_1 \in D_1, \LL, x_n \in D_n} \vdash \p{C'_2}{x_1 \in D''_1, \LL, x_n \in D''_n}$,
\II

\NI
and where $C'_1$ is the result of
restricting the constraint in $C_1$ to $D'_1, \LL,  D'_n$ and similarly for $C'_2$.
\II


In words: assuming we reduced the domains of each disjunct separately,
we can reduce the domains of the disjunctive constraint to the
respective unions of the reduced domains. 
As an example consider the constraint 
\[ 
\p{|x-y| = 1}{x \in [4..10], y \in [2..7]}.
\]
We can view $|x-y| = 1$ as the disjunctive constraint $(x-y = 1) \Orr (y-x=1)$.
In the presence of the {\it ARC CONSISTENCY} rules 1 and 2 rules we have then
\[
\p{x-y = 1}{x \in [4..10], y \in [2..7]} \vdash \p{x-y = 1}{x \in [4..8], y \in [3..7]}
\]
and
\[
\p{y-x = 1}{x \in [4..10], y \in [2..7]} \vdash \p{y-x = 1}{x \in [4..6], y \in [5..7]}.
\]
So using the \emph{CONSTRUCTIVE DISJUNCTION} rule we obtain
\[ 
\p{|x-y| = 1}{x \in [4..8], y \in [3..7]}.
\]

If each disjunct of a disjunctive constraint is a conjunction of constraints, the auxiliary derivations in the
side conditions can be longer than just one step.  Once the rules used in
these derivations are of an appropriate format, their applications can
be scheduled using one of the discussed generic iteration algorithms.  
Then the single application of the {\em CONSTRUCTIVE DISJUNCTION}  
rule consists in fact of two applications of the appropriate
iteration algorithm.

It is straightforward to check that if the auxiliary derivations
involve only monotonic domain reduction rules, then the {\em
  CONSTRUCTIVE DISJUNCTION} rule is itself monotonic.  So the {\sc
  Generic Iteration for Compound Domains} algorithm can be applied 
both within the side conditions of this rule and for scheduling this rule together
with other monotonic domain reduction rules that are used to deal with
other, non-disjunctive, constraints.

In this framework it is straightforward to formulate some strengthenings
of the constructive disjunction that lead to other modification of the
constraints $C_1$ and $C_2$ than $C'_1$ and $C'_2$.

\subsection*{Example 4: Propagation Rules}

These are rules that allow us to add new constraints.
Assuming a given set ${\cal A}$ of `allowed' constraints
we write such rules as
\II

\[
        \frac{{\cal B}}
             {{\cal C}}
\]
\II

\NI where ${\cal B}, {\cal C} \sse {\cal A}$.  

This rule states that 
in presence of all constraints in ${\cal B}$ the
constraints in ${\cal C}$ can be added, and is a shorthand for a deterministic
rule of the following form:
\II

\[
\frac{\p{{\cal B}}{x_1 \in D_1, \LL, x_n \in D_n}}
     {\p{{\cal B, C}}{x_1 \in D_1, \LL, x_n \in D_n}}
\]

An example of such a rule is the transitivity rule:
\II

\[
\frac{x<y, y<z}
     {x<z}
\]
\II

\NI
that refers to a linear ordering $<$ on the underlying domain
(for example natural numbers).

In what follows we focus on another example of propagation rules,
\oldbfe{membership rules}. They have the following form:
\[
\frac
{y_1 \in S_1, \LL, y_k \in S_k}
{z_1 \neq a_1, \LL, z_m \neq a_m}
\]
where $y_i \in S_i$ and $z_j \neq a_j$ are unary constraints with the obvious meaning.
\III

Below we write such a rule as:
\[
        y_1 \in S_1, \LL, y_k \in S_k \;\myra\; z_1 \neq a_1, \LL, z_m \neq a_m.
\]
The intuitive meaning of this rule is: if for all $i \in [1..k]$ the domain of
each $y_i$ is a subset of $S_i$, then for all $j \in [1..m]$ remove the
element $a_j$ from the domain of $z_j$.

The membership rules allow us to reason about constraints given
explicitly in a form of a table.  As an example consider the three
valued logic of Kleene. Let us focus on the conjunction constraint
\texttt{and3}$(x,y,z)$ defined by the following table:

\[
\begin{array}{|c|ccc|} \hline
               & $t$ & $f$ & $u$ \\ \hline
$t$            & $t$ & $f$ & $u$ \\
$f$            & $f$ & $f$ & $f$ \\
$u$            &  $u$ & $f$ & $u$ \\
\hline
\end{array}
\]  
That is, \texttt{and3} consists of 9 triples. Then the membership rule $y \in \C{u,f}
\myra z \neq t$, or more precisely the rule
\[
\frac
{\p{\texttt{and3}(x,y,z), y \in \C{u,f}}{x \in D_x, y \in D_y, z \in D_z}}
{\p{\texttt{and3}(x,y,z), y \in \C{u,f}, z \neq t}{x \in D_x, y \in D_y, z \in D_z}}
\]
is equivalence preserving.
This rule states that if $y$ is either $u$ or
$f$, then $t$ can be removed from the domain of $z$.  

We call a membership rule is \emph{minimal} if it is equivalence preserving
and its
conclusions cannot be established by either removing from its premise
a variable or by expanding a variable range.  For example, the above
rule $ y \in \C{u,f} \myra z \neq t $ is minimal, while neither $ x
\in \C{u}, y \in \C{u,f} \myra z \neq t $ nor $y \in \C{u} \myra z
\neq t $ is.  In the case of the \texttt{and3} constraint there are 18
minimal membership rules.

To clarify the nature of the membership rules let us mention that, as
shown in \cite{Apt00b}, in the case of two-valued logic the
corresponding set of minimal membership rules entails a form of
constraint propagation that is equivalent to the unit propagation, a
well-known form of resolution for propositional logic.  So the
membership rules can be seen as a generalization of the unit
propagation to the explicitly given constraints, in
particular to the case of many valued logics.

Membership rules can be alternatively viewed as a special class of
monotonic domain reductions rules in which the domain of each $z_i$
variable is modified by removing $a_i$ from it. So we can schedule
these rules using the {\sc Generic Iteration for Compound Domains}
algorithm.  

However, the propagation rules, so in particular the membership rules,
satisfy an important property that allows us to schedule them using a
more efficient, fine-tuned, scheduler. We call this property
\oldbfe{stability}.  It states that in each derivation the rule needs
to be applied at most once: if it is applied, then it does not need to
be applied again.  So during the computation the applied rules that
are stable can be \emph{permanently removed} from the initial rule set.
The resulting scheduler for the membership rules and its further
optimizations are discussed in \cite{BA05}.  

\section{Low Level}

The low level allows us to focus on matters that go beyond the issue
of rule scheduling.  At this level we can address matters concerned
with further optimization of the constraint propagation
algorithms. Various improvements of the \texttt{AC-3} algorithm that
are concerned with specific choices of the data structures used belong
here but cannot be explained by focusing the discussion 
on the corresponding \emph{ARC CONSISTENCY 1} and \emph{2} rules.   

On the other hand some other optimization issues can be explained in
proof-theoretic terms. In what follows we focus on the membership
rules for which we worked out the details.
These rules allow us to implement constraint propagation
for explicitly given constraints. We explained above that they can be
scheduled using a fine-tuned scheduler.  However, even when an
explicitly given constraint is small, the number of minimal membership
rules can be large and it is not easy to find them all. 

So a need arises to
generate such rules automatically. This is what we did in \cite{AM01}.
We also proved there that the resulting form of constraint propagation is equivalent
to \oldbfe{hyper-arc consistency}, a natural generalization of arc
consistency to $n$-ary constraints introduced in \mycite{MM88}.

A further improvement can be achieved by removing some rules
\emph{before} scheduling them. This idea was pursued
in \cite{BA05}.  Given a set of monotonic domain
reduction rules ${\cal R}$ we say that a rule $r$ is
\oldbfe{redundant} if for each initial CSP ${\cal P}$
the unique outcome of a stabilizing derivation
(guaranteed by Note \ref{note:monotonic}) is the same with $r$ removed
from ${\cal R}$.
In general, the iterated removal of redundant rules does not yield
a unique outcome but in the case of the membership rules
some useful heuristics can be used to appropriately
schedule the candidate rules for removal.

We can summarize the improvements concerned with the membership rules as follows:

\begin{itemize}
\item For explicitly given constraints all minimal membership rules can be automatically
generated.

\item Subsequently redundant rules can be removed.

\item A fine-tuned scheduler can be used to schedule the remaining rules.

\item This scheduler allows us to remove permanently some rules which is
useful during the top-down search.

\end{itemize}

To illustrate these matters consider the 11-valued
\texttt{and11} constraint used in the automatic test pattern
generation (ATPG) systems.  There are in total 4656 minimal membership
rules. After removing the redundant rules only 393 remain.  This leads
to substantial gains in computing.  To give an idea of the scale of
the improvement here are the computation times in seconds for three schedulers
used to find all solutions to a CSP consisting of the \texttt{and11}
constraint and solved using a random variable selection, domain ordering
and domain splitting:

\begin{table}[htbp]
  \centering
  \begin{tabular}{|l|l|l|l|}
\hline 
                     & Fine-tuned & Generic    & CHR \\\hline
all rules            & 1874       &  3321   & 7615 \\
non-redundant rules  & 157        & 316     & 543 \\
\hline 
  \end{tabular}
  \label{tab:and11}
\end{table}

\NI
\texttt{CHR} stands for the standard \texttt{CHR} scheduler normally
used to schedule such rules.  ({\tt CHR} is a high-level language
extension of logic programming used to write user-defined constraints,
for an overview see \cite{FruehwirthJLP98}.)
So using this approach a 50 fold improvement in computation time was achieved.
In general, we noted that the larger the constraint the larger the 
gain in computing achieved by the above approach.

\section{Conclusions}

In this paper we assessed the crucial features of constraint
programming (CP) by means of a proof-theoretic perspective.  To this
end we identified three levels of abstraction.  At each level proof rules
and derivations played a crucial role.  At the highest level they
allowed us to clarify the relation between CP and the computation as
deduction paradigm.  At the middle level we discussed
efficient schedulers for specific classes of rules.
Finally, at the lowest
level we explained how specific rules can be automatically
generated, optimized and scheduled in a customized way.  

This presentation of CP suggests that it has close links with 
the rule-based programming. And indeed, 
several realizations of constraint programming
through some form of rule-based programming exist.
For example, constraint logic programs are sets of rules, so
constraint logic programming can be naturally seen as an instance of
rule-based programming.  Further, the already mentioned {\tt CHR}
language is a rule-based language, though it does not have the full
capabilities of constraint programming.  In practise, \texttt{CHR} is
available as a library of a constraint programming system, for example
\eclipse{} (see \cite{eclipse05})
or SICStus Prolog (see \cite{sicstus05}).  In turn, {\sf ELAN}, see \cite{elan05}, is a
rule-based programming language that can be naturally used to explain
various aspects of constraint programming, see for example
\cite{ELANconstraints} and \cite{Cas98}.

In our presentation we abstracted from specific constraint programming
languages and their realizations and analyzed instead the principles
of the corresponding programming style. This allowed us to isolate the
essential features of constraint programming by focusing on proof
rules, derivations and schedulers.  This account of constraint
programming draws on our work on the subject carried out in the past
seven years.  In particular, the high level view was introduced in
\cite{Apt98a}. In turn, the middle level summarizes our work reported
in \cite{Apt99b,Apt00a}.  Both levels are discussed in more
detail in \cite{Apt03}.  Finally, the account of propagation rules and
of low level draws on \cite{AM01,BA05}.

This work was pursued by others. Here are some representative references.
Concerning the middle level, \cite{MR99} showed that the framework of
Section \ref{sec:middle} allows us to parallelize constraint
propagation algorithms in a simple and uniform way, while
\cite{BGR00a} showed how to use it to derive constraint propagation
algorithms for soft constraints.
In turn, \cite{Gen00a} explained other arc consistency algorithms by slightly
extending this framework.

Concerning the lowest level, \cite{RingeissenMonfroy:LNAI:2000}
considered rules in which parameters (i.e., unspecified constants) are
allowed. This led to a decrease in the number of generated rules.  In
turn, \cite{AR04} presented an algorithm that generates more general
and more expressive rules, for example with variable equalities in the
conclusion. Finally, \cite{AR05} considered the problem of generating
the rules for constraints defined intensionally over infinite domains.

\section*{Acknowledgments}

The work discussed here draws partly on a joint research carried out
with Sebastian Brand and with Eric Monfroy.
In particular, they realized the implementations discussed in the section
on the low level.
We also acknowledge useful comments of the referees.

\bibliographystyle{plain}

\bibliography{/ufs/apt/bib/apt,/ufs/apt/bib/sin02,/ufs/apt/bib/99,/ufs/apt/bib/clp1,/ufs/apt/esprit/esprit,/ufs/apt/bib/clp2}

\end{document}